\documentclass[12pt, preprint]{aastex}

\begin{document}

\title{Massive and Red Objects predicted by a semianalytical model of
galaxy formation} 

\author{X. Kang$^{1,2}$, Y. P. Jing$^{1}$, J. Silk$^{2}$}

\affil{$^1$ Shanghai Astronomical Observatory, Nandan Road 80, Shanghai, China}
\affil{$^2$ Astrophysics, University of Oxford, Denys Wilkinson
Building, Keble Road, Oxford OX1 3RH, UK}

\affil{e-mail: kangx@astro.ox.ac.uk}

\begin{abstract}
We study whether hierarchical galaxy formation in a concordance
$\Lambda$CDM universe can produce enough massive and red galaxies
compared to the observations. We implement a semi-analytical model 
in which the central black holes gain their mass during major
mergers of galaxies and the energy feedback from active galaxy nuclei
(AGN) suppresses the gas cooling in their host halos. The energy
feedback from AGN acts effectively only in massive galaxies when
supermassive black holes have been formed in the central
bulges. Compared with previous models without black hole formation,
our model predicts more massive and luminous galaxies at high
redshift, agreeing with the observations of K20 up to $z\sim 3$. Also
the predicted stellar mass density from massive galaxies agrees with
the observations of GDDS.  Because of the energy feedback from AGN,
the formation of new stars is stopped in massive galaxies with the
termination of gas cooling and these galaxies soon become red with
color $R-K>$5 (Vega magnitude) , comparable to the Extremely Red Objects (EROs) observed
at redshift $z\sim$1-2. Still the predicted number density of very
EROs is lower than observed at $z\sim 2$, and it may be related to
inadequate descriptions of dust extinction, star formation history and
AGN feedback in those luminous galaxies.

\end{abstract}


\section{Introduction}
There are many recent observations of high-redshift galaxies that
probe the star formation history of the Universe. The finding of many
massive galaxies, especially massive Extreme Red Objects (EROs), at
high redshift is particularly interesting. These observations show
that some EROs are passive ellipticals, and were already in place at
redshift z$\sim 2$.  It is usually argued that in a Cold Dark Matter
(CDM) universe, structures form via a hierarchical formation process
in which small galaxies form first at early times, and massive
galaxies form later through the continuous mergers of the smaller
systems. With representative semi-analytical models (SAMs; Kauffmann
et al. 1999, Somerville \& Primack 1999, Cole et al. 2000), it was
found that in the concordance $\Lambda$CDM universe, it is difficult
to produce enough massive and red galaxies that look like those
observed(e.g. Cimatti et al. 2002a, Glazebrook et al. 2004). On the
other hand, the existence of the observed massive galaxies at high
redshift is not necessarily in conflict with the concordance
$\Lambda$CDM model, because the conversion of just ten percent of
baryons in dark matter halos of mass $M >10^{13}M_{\odot}$ to stars is
sufficient to produce the number of observed massive galaxies
(Somerville 2004a).

Many authors have studied the formation of these massive, red objects
using SAMs or Smoothed Particle Hydrodynamics (SPH) simulations. It
was shown that the SAMs (Kauffmann et al. 1999, Somerville \& Primack
1999, Cole et al. 2000) cannot produce enough massive/red objects at
redshift $z>1$ (e.g. Firth et al 2002, Somerville et al. 2004b, Daddi
et al. 2005). The SPH simulations (e.g. Nagamine et al. 2004, 2005)
have succeeded in producing massive and red galaxies at high redshift,
but at the cost of introducing more uncertainties. First, it is
unknown if these SPH simulations can produce the local galaxy
luminosity function. It seems that these simulations produce too many
bright galaxies at $z=0$ (Nagamine et al. 2004). Secondly, Nagamine 
et al. (2005) used a high dust extinction for the entire galaxy population, 
but the observations show that some EROs are passive
ellipticals with little dust extinction (Cimatti et al. 2002b).

The main reason that the SAMs fail to produce enough massive and
luminous galaxies at high redshift is that the gas cooling and star
formation in early massive halos is over-suppressed. In previous SAMs,
the gas cooling in massive halos is switched off in order not to
produce more luminous central galaxies than observed at redshift
$z=0$. The suppression of gas cooling is also motivated by the X-ray
observations that massive cooling flows are not observed in groups and
clusters (e.g. Peterson et al. 2003). But as the consequence, the gas
cooling may be over-suppressed at high redshift if a simplified
prescription is used for the cooling cutoff.  For example, in the
Munich group model and also in Kang et al. (2005), the gas cooling is
shut off by hand in halos with the virial velocity greater than
$350km/s$. Since the halo mass is much lower at high redshift than at
the present for a given virial velocity, the gas cooling is suppressed
in this model for halos with the virial mass greater than 2.5$\times
10^{12}M_{\odot}$ at z $=$ 3. This artificial cooling switch-off seems
to be the main reason that these models do not produce as many massive
galaxies as observed.

In this paper, we implement a new model in which the energy from AGN
is used to suppress the cooling of hot gas in halos.  Following
Kauffmann \& Haehnelt (2000) we use a simple model wherein black holes
gain most of their mass during major mergers. Our implementation of
the feedback from AGN is very similar to that used recently by Croton
et al. (2006) and Bower et al. (2005), and resembles a combination of
their models. In our model, the total energy from the AGN is
proportional to the Eddington luminosity of the central black hole and
the efficiency of reheating the gas is proportional to a power of the
virial velocity of the galaxy. Then the energy compensates for the
radiative energy of the cooling gas, and the actual cooling rate is
determined by the ratio between the two energies. The cooling is
totally suppressed if the energy from AGN is larger than the energy
radiated by the cooling gas.  Compared with the previous model used by
Kang et al. (2005) with an artificial cut-off of the gas cooling in
the halos with the virial velocity larger than $350km/s$, the gas
cooling and AGN feedback in the new model are treated in a more
self-consistent way.  The $M_{\rm bh}$-$\sigma$ relation of black hole
mass $M_{\rm bh}$ and the bulge velocity dispersion $\sigma$ implies
that massive black holes are present only in massive spheroids. In our
present model, the energy feedback from AGN indeed is efficient in
galaxies with a massive spheroid. We also require that the star
formation rate in quiescent disks is reduced at high redshift as
motivated by the observed evolution of cosmological cold gas content
with redshift (Keres et al. 2005); thus the gas-rich mergers result in
earlier formation of supermassive black holes in massive central
bulges. Once the energy feedback is enough to suppress the gas
cooling, the termination of new star formation will soon make the
galaxies red. We will compare the model prediction of the number density
of luminous galaxies with the K20 survey, and find that good agreement
holds up to z$\sim$3, beyond which there is little observational
data. Compared with previous SAMs, our present model can also produce
some very red ($R-K>5$, magnitudes are given in the Vega system unless
otherwise stated) passive ellipticals which are observed by the Great
Observatories Origins Deep Survey (GOODS) at z$\sim 1-2$.

We arrange our paper as follows. In section 2, we briefly introduce
our new model with AGN feedback and compare our model predictions with
the local galaxy population. In section 3, we give the model
predictions and compare them with the observations at high redshift.
Finally, we discuss our results and conclude our work in section 4.

\section{Model}
The SAM that we use here was described in detail by Kang et
al. (2005) who studied the local galaxy population. The merger tree is
constructed based on a high-resolution N-body simulation (Jing \& Suto
2002) of 512$^{3}$ particles in a box of 100$h^{-1}{\rm Mpc}$. The
cosmological parameters adopted there are $\Omega_{m} = 0.3$,
$\Omega_{\Lambda} = 0.7$, $h=0.7$, $\sigma_{8} = 0.9$.  Here we still
use this simulation, but the SAM model is modified in two ways.

1. We adopt a star formation efficiency $\alpha \sim (1+z)^{-1}$ in a
quiescent disk that was shown to give a better match with the
evolution of cosmological cold gas content with redshift (Kauffmann \&
Haehnelt 2000, P$\acute{\rm e}$roux et al. 2003, Keres et al. 2005).
In the recent model of Durham group (Baugh et al. 2005, Bower et
al. 2005), they adopt a constant star formation timescale for the
disk. The star formation timescale used in our model is the dynamical
time of the disk which scales with redshift as $(1+z)^{-1.5}$. So the
star formation rate ($\dot{M_{\ast}}=\alpha M_{cold}/t_{dyn}$) of our
model differs from that of the Durham model only slightly. Note that
the relatively lower star formation rate in quiescent disks leaves
more cold gas which helps to produce massive black holes during galaxy
mergers at high redshift.

2. We include a model for the growth of black holes and for the energy
feedback from AGN to suppress the gas cooling. As the $M_{\rm
  bh}$-$\sigma$ relation indicates that the central black holes grow
with the growth of the spheroid components, it is plausible that the
black holes get their mass through major mergers. But it is far from
clear about the exact way that the black holes accrete the surrounding
material. Here following Kauffmann \& Haehnelt (2000), we use a
simple parameterised form to describe the cold gas accreted by the
black hole during a major merger,
\begin{equation}
\Delta M_{bh} = F_{acc} \frac {M_{cold}} {1+(280km/s/V_{vir})^{2}}
\end{equation}
where $M_{cold}$ is the total cold gas in merging galaxies, and
$V_{vir}$ is the virial velocity of the post-merger host halo. We
normalize the parameter $F_{acc}$ by best matching the observed
$M_{bulge}-M_{bh}$ relation at z=0 (H\"aring \& Rix 2004).  During the
gas accretion by black holes, part of the gravitational energy will be
converted into radiations which in turn will heat the surrounding cold
gas. But it is again unclear in a quantitative way about how much the
radiation is produced and how efficiently the cold gas is re-heated.
Croton et al. (2006) use a simple phenomenological model to describe
the accretion rate which depends on the hot gas fraction and circular
velocity of the halo, but the efficiency of heating the gas by AGN are
the same in all halos of different mass. Sijacki \& Springel (2006)
have shown that heating efficiency from a AGN bubble is lower in low
mass halos. Here we simply assume that the energy from the central AGN
is proportional to the Eddington luminosity $L_{edn}$ and the heating
efficiency is proportional to a power of the virial velocity of
the host halo. Thus the heating rate ejected into the gas is taken as,
\begin{equation}
L_{reheat}=F_{0}(V_{vir}/V_{\star})^{n}L_{edn}\,.
\end{equation}
If we denote the cooling rate in a halo of gas temperature $T$ by
$\dot{M}_{0,cool}$ in the case of no AGN feedback, then the cooling
rate $\dot{M}_{cool}$ in the presence of AGN feedback is:
\begin{equation}
\frac {\dot{M}_{cool}} {\dot{M}_{0,cool}} = 1 - \frac {L_{reheat}} {\frac {3} {4}\dot{M}_{0,cool}V_{vir}^{2}}.
\end{equation}
If the heating rate from AGN $L_{reheat}/\frac{3}{4}V_{vir}^2$ is
larger than the radiative cooling rate ${\dot{M}_{0,cool}}$, the gas
cooling is totally suppressed.  We normalize the parameters $F_{0}$,
$V_{\star}$ to get a good match to the galaxy luminosity function at
z=0. In our model we obtain $F_{0}=2\times 10^{-5}$ and
$V_{\star}=200km/s$ and $n=4$.

\begin{figure}
\plotone{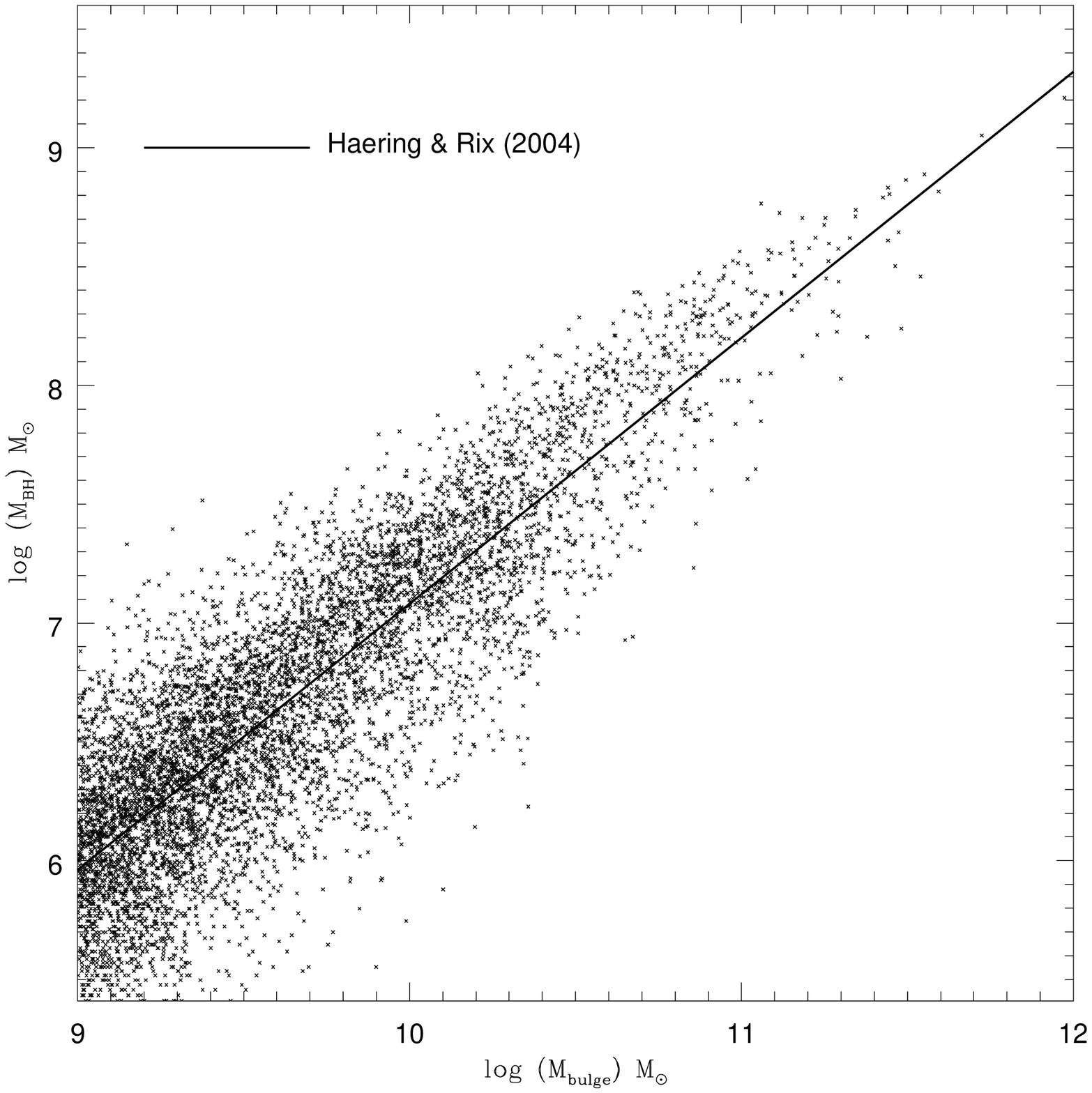}
\caption{The relation between the mass of the bulge and the black hole for local galaxies. The 
crosses are the model galaxies and the solid line is the best fit to the observations (H\"aring \& Rix 2004).}
\label{fig:Bh_Bulge}
\end{figure}

\begin{figure}
\plotone{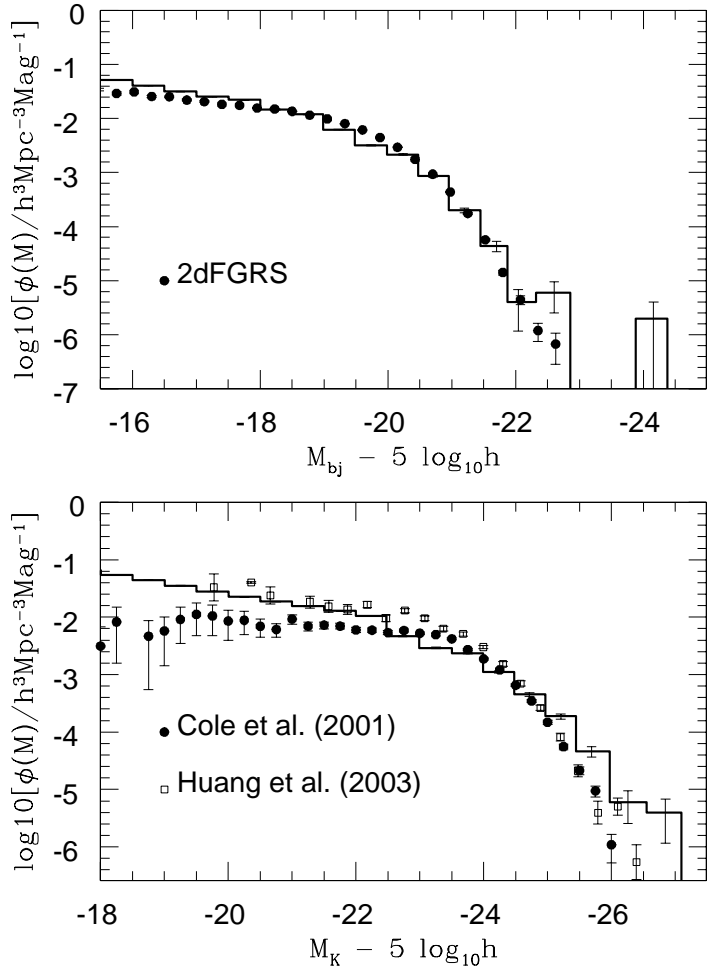}
\caption{The local galaxy luminosity function. Upper panel: the
B$_{j}$ band luminosity function, and the data points are from the
2dFGRS (Norberg et al. 2001). Lower panel: the K band luminosity
function, and the data points are from observations of Cole et
al. (2001) and Huang et al. (2003). In both panels, the histograms
associated with Poisson errors are the model predictions.}
\label{fig:LF_z0}
\end{figure}

In Fig.~\ref{fig:Bh_Bulge} we plot the relation between the bulge mass
and the black hole mass. The data points show for the model galaxies
and the solid line the best fit to the observations by H$\ddot{\rm
a}$ring \& Rix (2004). Here $F_{acc}$ is taken to be $0.01$. It is
seen that a simple model of black hole growth with a free parameter
can reproduce the observed $M_{bulge}-M_{bh}$ relation.  After the
black hole mass is normalized, we then tune the parameters in equation
2 to get good fits to the local galaxy luminosity functions. In
Fig.~\ref{fig:LF_z0} we show the luminosity function at B$_{j}$ and K
bands. The upper panel shows a comparison with the 2dFGRS at B$_{j}$
band.  The solid circles show the observational data of 2dFGRS, and
the thick solid histogram associated with Poisson errors is our model
prediction.

The lower panel shows the comparison at K band where the circles are
from Cole et al. (2001) and squares are the observations by Huang et
al. (2003).  We find that the new model can produce the local galaxy
luminosity functions at blue and near-IR bands which are respectively
sensitive to the current star formation rate and the total stellar
mass in the galaxies. It has been shown (Croton et al. 2006, Bower et al. 2005)
 that without an effective energy feedback, the predicted
luminosity functions at the bright end are too flat with many more
luminous galaxies predicted than observed. Note that here our model
predictions at high luminosity ends are still slightly higher than
observed. This might point to the fact that a more detailed model is
needed for AGN heating in massive halos which we will address in
future work.

\section{Results at high redshift}

\begin{figure}
\plotone{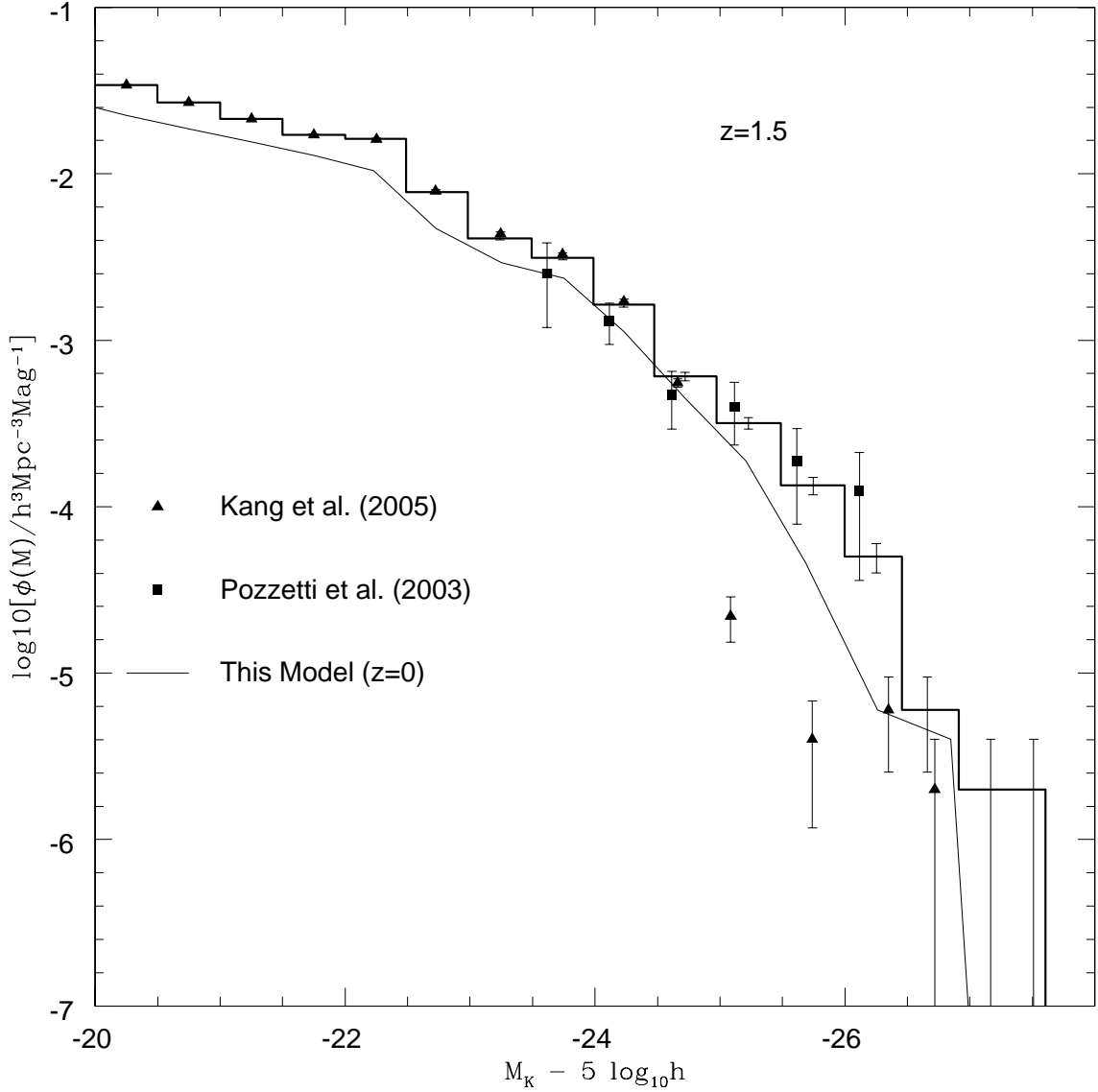} 
\caption{The rest-frame K band luminosity function at $z=1.5$. The
data points with errorbars are the observations of K20 (Pozzetti et
al. 2003). The histograms are our new model with AGN feedback, and the
triangles are the prediction by Kang et al. (2005) where they used an
artificial cooling cut for halos with $V_{vir}>350km/s$. The solid
line is the local K band luminosity function from our new model, shown
also as the histogram in the lower panel of Fig2.}
\label{fig:K20_LF}
\end{figure}

As discussed in Section 1, the gas cooling in our new model is not
suppressed artificially but by heating due to the energy injected from
AGN in the galaxy center. So compared to previous SAMs without
AGN, the gas cooling and star formation can continues until a massive
spheroid forms at the galaxy center. It is expected that this model
can produce more massive and luminous galaxies at high redshift. In
Fig.~\ref{fig:K20_LF} we show the predicted rest-frame K band
luminosity function at z$\sim 1.5$. The squares with error bars are
the observational results from K20 (Pozzetti et al. 2003).  The solid
circles are the predictions by the new model and the triangles show
the results predicted by Kang et al. (2005) where
they adopted a artificial shut off of gas cooling in galaxies with
$V_{vir}>350km/s$. We also re-plot the results of K band luminosity
function at z=0 by the solid line, taken from from lower panel of
Fig.\ref{fig:LF_z0}.  It is clearly seen from the plot that the new
model produces more massive galaxies and the agreement with the
observations is very good. Also note that the good agreement holds for
faint galaxies as well, whereas it was reported previously that SAM
models produce more faint galaxies than observed (Pozzetti et
al. 2003).

\begin{figure}
\plotone{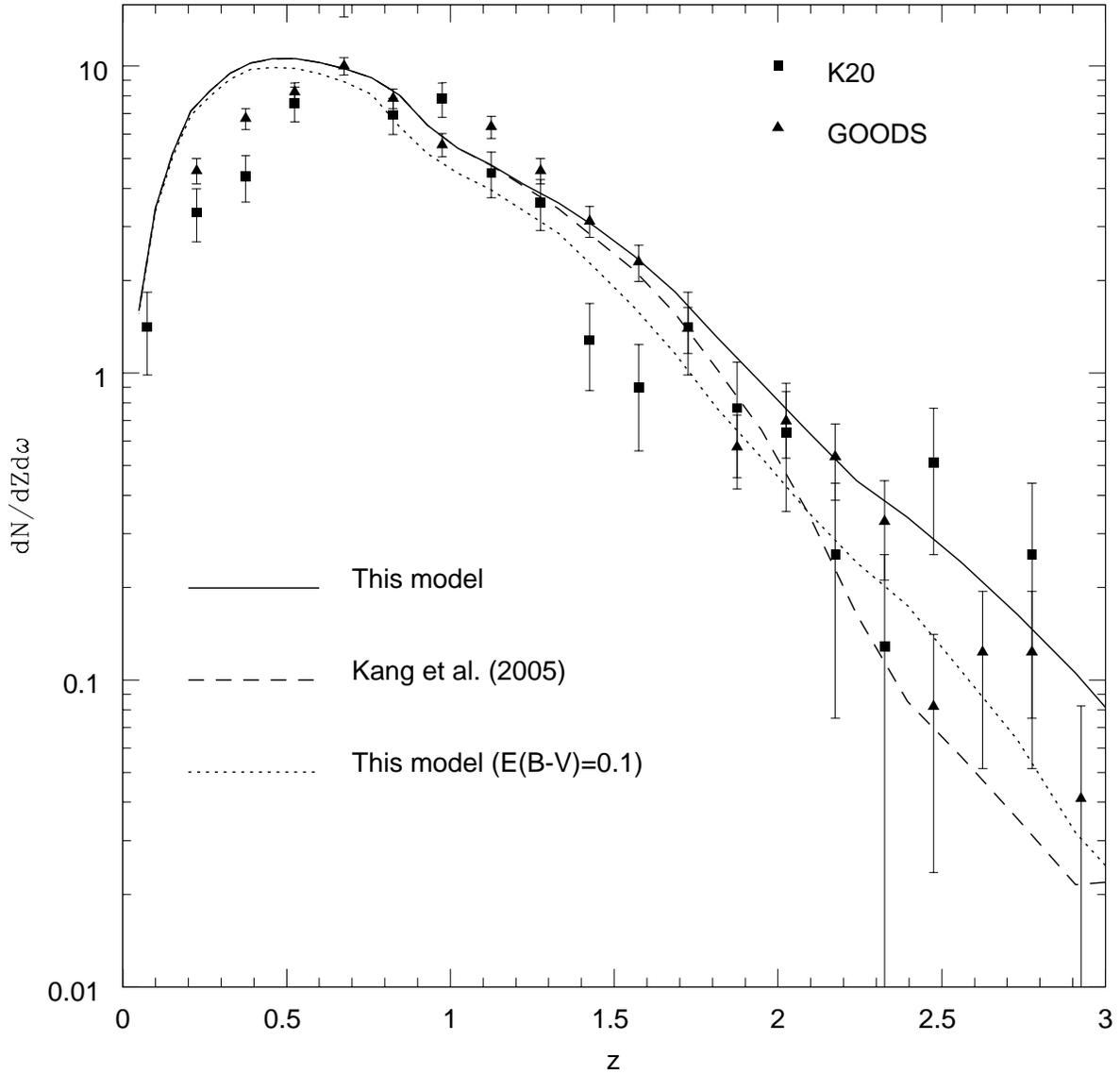} 
\caption{The redshift surface density distribution of $K<20$ galaxies. Data points show the observations,
solid and dashed lines are the model results without dust extinctions, and the dotted line is 
for a simple reddening law of Calzetti et al. (2000) with $E(B-V)=0.1$.}
\label{fig:GOODS_num}
\end{figure}

Another test, firstly proposed by Kauffmann \& Charlot (1998), is the
evolution of the surface number density of galaxies at a fixed
limiting magnitude, which also widely used to constrain the
models. There are plenty of data from GOODS that are already publicly
available (Giavalisco et al. 2004). In Fig.~\ref{fig:GOODS_num} we
show the predicted redshift surface number density of galaxies with K$<20$.
The square points show the results of K20 and triangles are the data
from GOODS.  The new model predictions are shown as the solid line,
and the dashed line shows the prediction by the model of Kang et
al. (2005). Here we find that compared with Somerville et al. (2004b)
who predicted much fewer luminous galaxies at $z>1.5$, the agreement
between our model and the observations holds much better up to z$\sim
3$. Here we also show how dust extinction will change the result. The
dotted line is the new model with the simple dust extinction model of
Calzetti et al. (2000) with $E(B-V)=0.1$. Clearly dust extinction has
no significant effect on the predicted number of galaxies in the
observed-frame K band up to z=3.

Though the predicted numbers of luminous galaxies agree with the
observations, it would be interesting to check the predicted color
distributions. The color is dependent on the star formation history
and on the dust extinction. At high redshift the galaxy mergers are
very frequent and the dust extinction is significant in these
starburst galaxies, but no reliable model of dust extinction is
available for such galaxies.  Observations show that at z$\sim 1-2$
the EROs have contributions both from passive ellipticals with little
dust and from dust-enshrouded starburst galaxies (Cimatti et al. 2002b, Cimatti et al. 2003, 
Yan \& Thompson 2003, Yan et al. 2004, Moustakas et al. 2004). Because there are
significant uncertainties in the dust extinction modelling for the
starburst galaxies, we think that the predicted number density of
passive ellipticals should set a more meaningful constraint on the
galaxy formation model. Here we take a simple model of dust
extinction. We classify the galaxies with starbursts produced during
the major mergers in the past 0.1Gyr as young starburst galaxies
and those otherwise as passive galaxies.  We then use the Calzetti et
al. (2000) reddening law to model the dust extinction effect on the
galaxy color. The amount of dust in passive and young starburst
galaxies is  difficult to assess, and here we simply assume a small
reddening $E(B-V)=0.05$ for the passive galaxies. The dust extent in
young starburst galaxy is expected to be high.  Observations of
EROs show that some extremely red galaxies have heavy dust
extinction with $E(B-V)=0.4$. But the average extinction should be
lower. Here we assume a Gaussian distribution of $E(B-V)$ with a mean of
0.1 and a dispersion of 0.05 for the young starburst galaxies. Our
main motivation is to see if a simple dust reddening model can produce
the main features of the observed color distribution.

\begin{figure}
\plotone{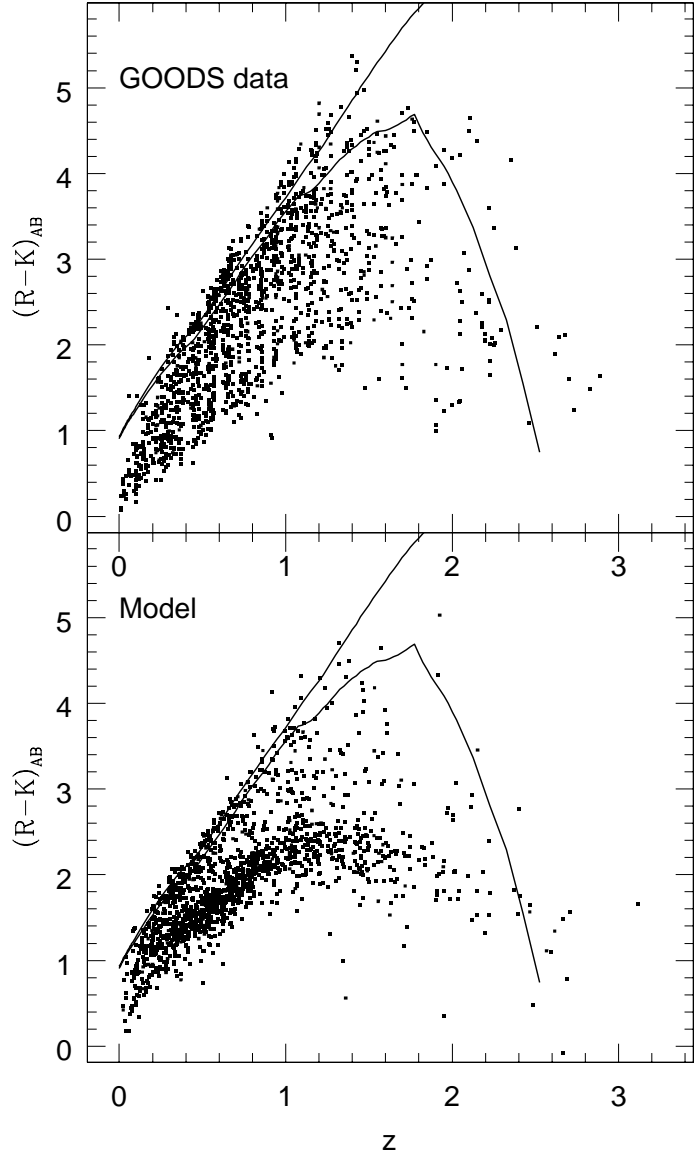} 
\caption{$R-K$ color distributions in the observed frame. Upper panel:
the GOODS data from Fig.2 of Somerville et al. (2004a). The lower
panel: the model galaxies with a simple dust extinction.}
\label{fig:GOODS_color}
\end{figure}

In Fig.~\ref{fig:GOODS_color} we show the observed $R-K$ (both in the
AB magnitude system, $(R-K)_{AB} \simeq (R-K)_{Vega}-1.65$) color distribution with a comparison with the
data which are from the GOODS Southern field in an area of 160
arcmin$^{2}$ (Somerville et al. 2004b). The upper panel shows the GOODS
data, which is from Figure 2 of Somerville et al. (2004b). The model
galaxies are selected using the magnitude cut and are normalized to
the same area of 160 arcmin$^{2}$. The total number of galaxies
selected in our model is 1595 which is $6\%$ higher than the GOODS
data points used here. The lower panel shows the model predictions.
In each panel we also show the evolution track of single burst stellar
populations with solar metallicity, the Salpeter IMF, and the ages (at
$z=0$) of 13.35 and 11.7 Gyrs (i.e. $z_{f}=26, 2.6$) based on the
model of BC03 (Bruzual \& Charlot 2003). From the figure, our model
can reproduce the main features of the observed galaxies: 1) many
extremely red galaxies ($R-K>4$) at $z>1$; 2) the bimodal color
distribution, red passive and young starburst galaxies at
$z>1.5$. Still there are some discrepancies. The predicted numbers of blue
galaxies are too prominent at z$<1.5$ and this might be due to the
inadequate treatment of star formation rate, stellar initial mass
function, or the dust extinction model. Also the predicted number of
extremely red galaxies with $(R-K)_{AB}>3.35$ at $z \sim 2$ is still
lower than observed.  In our model there are enough luminous galaxies
but insufficient number of very red galaxies, which means that the
star formation (at $\sim 2$) in the current model are still high. There are two
possible reasons for this discrepancy. First the star formation is
not strong enough in the past in our model, as we do not include any
star formation during minor mergers which are also frequent at early
times.  Second the energy from central AGN is not high enough to
suppress the hot gas cooling.  Observations have shown that there are
already massive black holes ($\sim 10^{9}M_{\odot}$) at $z \sim 6$
(Fan et al. 2001), so the growth of black holes in massive galaxies
might be much quicker at early time than in our model in which the
fraction of cold gas accreted by black hole is constant with time. We
will address this in a forthcoming paper (Kang et al. 2006).

\begin{figure}
\plotone{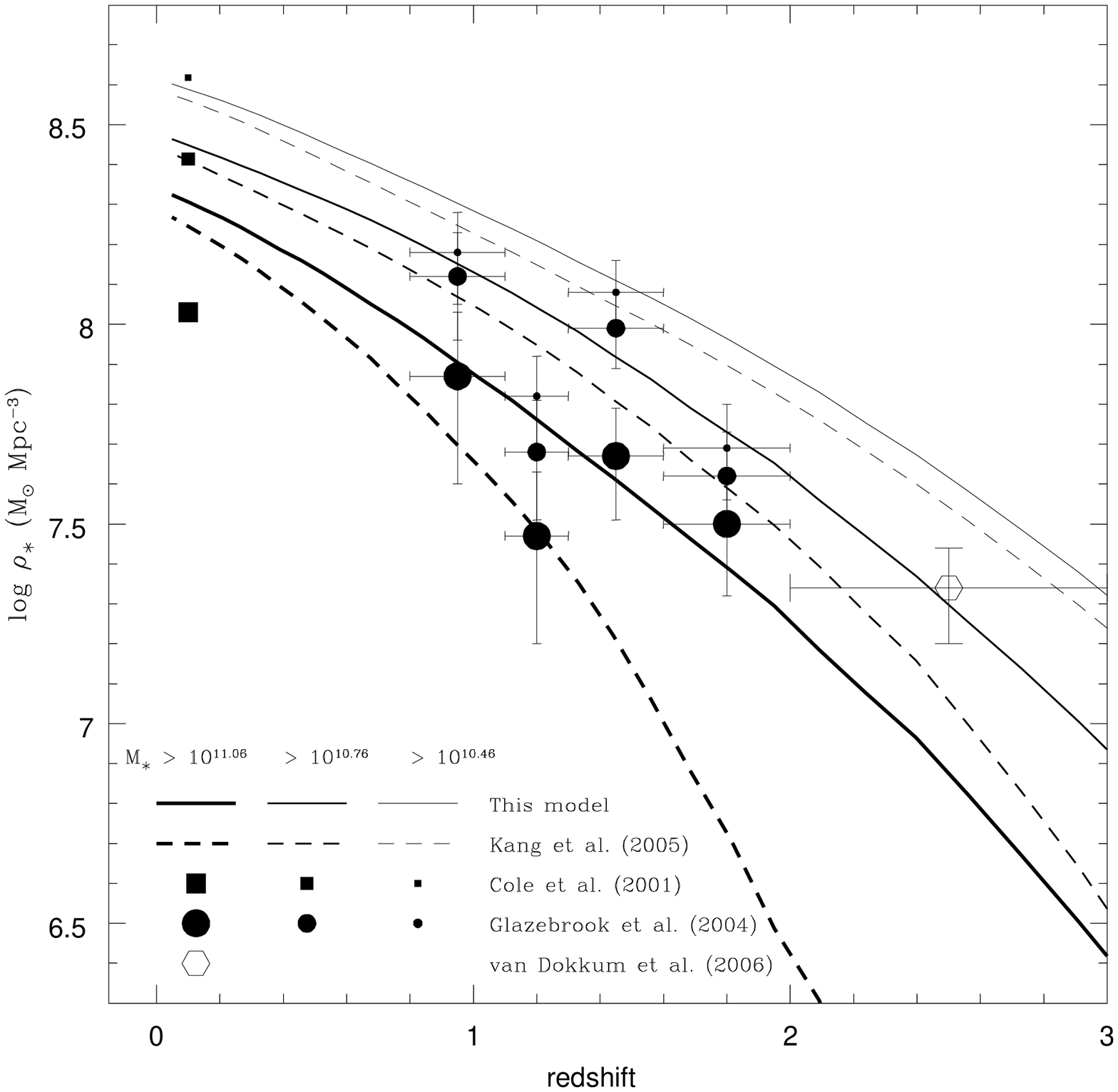} 
\caption{The stellar mass density as a function of redshift for
galaxies with stellar mass above certain threshold.  The mass
thresholds are indicated in the plot. The data points are from
Glazebrook et al. (2004), but have been transformed to the case of the
Salpeter IMF. The lines are the model results, see the text.}
\label{fig:stellar_GDDS}
\end{figure}

Glazebrook et al. (2004) used the Gemini Deep Deep Survey (GDDS) to
obtain the stellar mass distribution from $z \simeq$ 0.7 to 2. The
evolution of stellar mass density does place important constraints on
the formation model of massive spheroids. But due to the uncertainties
in fitting the multi-broad band colors of high redshift galaxies
including those of the IMF and dust extinction, the constraints are
weak. In Fig.~\ref{fig:stellar_GDDS}, we show the stellar mass
density of  galaxies with  stellar mass above certain
limits. The lines show the predicted stellar density in galaxies with
stellar mass in the range indicated in the plot. Black lines
are for this model and blue lines are from the model of Kang et
al. (2005) where they used an artificial cut of gas cooling in the halos
with $V_{vir} > 350 km/s$.  We can still see a good match between the
model and the data. Although it seems that the stellar mass density
with $M_{\star}> 10^{10.46}M_{\odot}$ is higher than the data points,
it agrees with the integral of the star formation rate (see figure 4
of Glazebrook et al. 2004). Note that galaxies with
$M_{\star}>10^{11}M_{\odot}$ are in the sharply declining tail of the
mass function, therefore a small uncertainty in the estimated stellar
mass can introduce a very large uncertainty in the number density. 
The hexagon in the plot shows the stellar mass density of massive galaxies with
$M_{\star}>10^{11}M_{\odot}$ recently obtained by van Dokkum et
al. (2006) making use of the deep multi-wavelength GOODS, FIRES and MUSYC surveys. 
It is seen from the black dashed lines that our
model prediction is slight lower than the data by a factor of 2. At
high redshift the cosmic variance is so large in the observed catalogs
(about $60\%$, Somerville et al. 2004c) that the discrepancy might not
be serious.

\section{Discussion}
Here we have implemented a new semi-analytical model in which the
energy from AGN suppresses hot gas cooling in massive
halos. The growth of black holes and bulges, and the gas cooling, are
determined in a self-consistent way. In our description, the AGN
feedback becomes efficient in massive galaxies after a massive black
hole is formed in the galaxy center. The AGN feedback model has drawn
much recent attentions. The main motivation is that in massive groups
and clusters cooling flows are not observed. There should be some
physical process to reheat the cooling region, and the energy from AGN
has been proposed as an effective source (e.g. B$\ddot{\rm o}$hringer
et al. 2002, Begelmen et al. 2002, Sijacki \& Springel 2006). At the
same time, the AGN feedback models have also been incorporated into
the SAMs recently and it has been shown that AGN feedback can produce
a break of the luminosity function at the bright end and produce the
color-magnitude relation observed in SDSS (Croton et al. 2006, Bower
et al. 2005). Our model of AGN feedback is very similar to theirs in
spirit, but the detailed prescription is different. In this paper
we use this model to address some issues about the number distribution
and color distribution of galaxies at high redshift. We compare the
model predictions with the K20 and GOODS surveys. Our conclusions are
as follows.
\begin{itemize}
\item The predicted number distribution of $K<20$ galaxies matches
well with that of the GOODS and K20 galaxies up to a redshift of z
$\sim$ 3;
\item The predicted color distribution is similar to that observed in
the surveys and many extremely red galaxies ($R-K_{AB}>4$) are
produced, which has  not been seen in previous models (Somerville et
al. 2004b).  At $z > 1.5$ the galaxy population already displays a
bimodal color distribution;
\item The predicted stellar mass density can marginally agree with the
GDDS observation even with the uncertainties in the IMFs;
\end{itemize}
These results demonstrate that it is not difficult to produce massive
and red galaxies at z $\sim$ 1-2 in the concordance CDM universe.  The
stellar mass in galaxy centers continues to grow until the energy from
central AGN is high enough to suppress the gas cooling. In our model
the black holes acquire most of their mass during major mergers, so
the AGN energy feedback is expected to be effective after the last
major merger which led to massive bulge formation at galactic centers.
In our model we can produce some of those passive ellipticals at
z$\sim 1-2$ with extremely red colors $(R-K)_{AB}>4$.

Many observations have shown that the star formation rate was higher
in massive galaxies at high redshift and these support the
"downsizing" formation scenario (Cowie et al. 1996). It is often argued that hierarchical galaxy
formation cannot reproduce the downsizing formation process. But
recent works (de Lucia et al. 2005, Bower et al. 2005, Scannapieco et
al. 2005) have shown that models with AGN feedback in the hierarchical
universe can reproduce the downsizing process in which the massive
galaxies forms earlier. In this paper, we also find that the predicted
luminous and massive galaxies are increased to the degree that is in
agreement with the observations, though the predicted number of red
galaxies may still be fewer than observed. Once more observations are
available on the dust extinction in these galaxies, the number density
and evolution of red passive ellipticals will put more stringent
constraints on the galaxy formation models. It is also possible that a
new ingredient is needed, such as the star formation induced by AGN
feedback prior to disruption of the cold gas supply (Silk 2005), in
order to make bulge formation more efficient and to account for the
chemical evolution of massive early-type galaxies.

\acknowledgments 
We thank Mashiro Nagashima for kindly providing the GOODS data, and
Manfred Georg Kitzbichler for the binned data of GOODS and K20. Xi
Kang acknowledge support from the Royal Society China Royal Fellowship
Fellowship scheme. This work is supported in part by
NSFC(No. 10373012, 10533030) and by Shanghai Key Projects in Basic
research (04jc14079, 05xd14019).

\newpage
\clearpage

\clearpage

\clearpage

\clearpage

\clearpage

\clearpage

\begin{thebibliography}{}
\bibitem[Baugh et al. 2005]{Baugh05}
Baugh, C. M., Lacey, C. G., Frenk, C. S., Granato, G. L., Silva, L., Bressan, A., Benson, A. J., \& Cole, S., 2005, MNRAS, 356, 1191
\bibitem[Begelman 2000]{begelman00}
Begelman, M. C., 2003, in ``Carnegie Observatories Astrophysics Series,
 Vol. 1: Coevolution of Black Hole and Galaxies, ``ed. L. C. Ho, 
Cambridge: Cambridge Univ. Press, astro-ph/0303040
\bibitem[Benson et al. 2003]{benson03}
Benson, A. J., Bower, R. G., Frenk, C. S., Lacey, C. G., Baugh, C. M., 
\& Cole, S., 2003, ApJ, 599, 38
\bibitem[Boehringer et al. 2002]{Boehringer02}
B$\ddot{\rm o}$hringer, H., Matsushita, K., Churazov, E., Ikebe, Y., \& Chen, Y., 2002, A\&A, 382, 804
\bibitem[Bower et al. 2005]{Bower05}
Bower, R. G., Benson, A. J., Malbon, R., Helly, J. C., Frenk, C. S., Baugh, C. M., Cole, S., \& Lacey, C. G., 2005, preprint (astro-ph/0511338)
\bibitem[BC03]{bruzual03}
Bruzual, G., \& Charlot, S., 2003, MNRAS, 344, 1000
\bibitem[Calzetti et al. 2000]{CLazetti00}
Calzetti, D., Armus, L., Bohlin, R. C., Kinney, A. L., Koornneef, J.,
\& Storchi-Bergmann, T., 2000, ApJ, 533, 682
\bibitem[Cimatti et al. 2002a]{cimatti02a}
Cimatti, A., et al., 2002a,A\&A, 391, L1
\bibitem[Cimatti et al. 2002b]{cimatti02b}
Cimatti, A., 2002b, A\&A, 381, L68
\bibitem[Cimatti et al. 2003]{cimatti03}
Cimatti, A., 2003, A\&A, 412, L1
\bibitem[Cole et al. 2000]{cole00}
Cole, S., Lacey, C. G., Baugh, C. M., \& Frenk, C. S., 2000, MNRAS, 319, 168
\bibitem[Cole et al. 2001]{cole01}
Cole, S., et al. 2001, MNRAS, 326, 255
\bibitem[Cowie et al. 1996]{cowie96}
Cowie, L. L., Songaila, A., Hu, E. M., \& Cohenm J. G., 1996, AJ, 112, 839 
\bibitem[Darren et al. 2006]{darren06}
Croton, D. J., Springel, V., White, S. D. M., De Lucia, G., Frenk, C. S., Gao, L, 
Jenkins, A., Kauffmann, G., Navarro, J. F., \& Yoshida, N., 2006, MNRAS, 365, 11
\bibitem[Daddi et al. 2005]{daddi05}
Daddi. E., et al. 2005, ApJ, 626, 680
\bibitem[De Lucia et al. 2005]{Delucia05}
De Lucia, G., Springel, V., White, S. D. M., Croton, D., \& Kauffmann, G., 2006, MNRAS, 366, 499
\bibitem[Di Matteo et al. 2005]{dimatteo05}
Di Matteo, T., Springel, V., \& Hernquist, L., 2005, Nature, 433, 604
\bibitem[Giavalisco et al. 2004]{giavalisco04}
Giavalisco, M., et al. 2004, ApJ, 600, L93
\bibitem[GLazebrook et al. 2004]{glazebrook04}
Glazebrook, K., et al. 2004, Nature, 430, 181
\bibitem[Fan et al. 2001]{fan01}
Fan, X., et al., 2001, AJ, 122, 2833 
\bibitem[Firth et al. 2002]{firth02}
Firth, A. E., et al. 2002, MNRAS, 332, 617
\bibitem[Haering \& Rix 2004]{haering04}
H\"aring, N., \& Rix, H. W., 2004, ApJ, 604L, 89
\bibitem[Huang et al. 2003]{huang03}
Huang, J. S., Glazebrook, K, Cowie, L. L., \& Tinney, C., 2003, ApJ, 584, 203
\bibitem[Jing \& Suto 2002]{jing02}
Jing, Y. P., \& Suto, Y., 2002, ApJ, 574, 538
\bibitem[Kang et al. 2005]{kang05}
Kang, X., Jing, Y. P., Mo, H. J, \& B$\ddot{\rm o}$orner, G., 2005, ApJ, 631, 21
\bibitem[Kauffmann \& Charlot 1998]{kauffmann98}
Kauffmann, G., \& Charlot, S., 1998, MNRAS, 297, 23
\bibitem[Kauffmann et al. 1999]{kauffmann99}
Kauffmann, G., Colberg, J. M., Diaferio, A., \& White, S. D. M., 1999, MNRAS, 303, 188
\bibitem[Kauffmann \& Haehnelt]{Kauffmann00}
Kauffmann, G., \& Haehnelt M, 2000, MNRAS, 311, 576
\bibitem[Keres et al. 2005]{keres05}
Keres, D., Katz, N., Weinberg, D. H., \& Dave, R. 2005, MNRAS, 363, 2
\bibitem[Moustakas et al. 2004]{moustakas04}
Moustakas, L. A., et al. 2004, ApJ, 600, L131
\bibitem[Nagamine et al. 2004]{nagamine04}
Nagamine, K., Springel, V., Hernquist, L., \& Machacek, M., 2004, MNRAS, 350, 385
\bibitem[Nagamine et al. 2005]{nagamine05}
Nagamine, K., Cen, R., Hernquist, L., Oxtriker, J. P., \& Springel, V., 2005, ApJ, 627, 608
\bibitem[Narayan 2001]{Naraayan01}
Narayan, R., \& Medvedev, M. V., 2001, ApJ, 562L, 129
\bibitem[Peroux et al. 2003]{peroux03}
P$\acute{\rm e}$roux, C., McMahon, R. G., Storrie-Lombardi, L. J., \& Irwin, M. J., 2003, MNRAS, 346, 1103
\bibitem[Peterson et al. 2003]{perterson03}
Peterson, J. R., Kahn, S. M., Paerels, F. B. S., et al. 2003, ApJ, 590, 207
\bibitem[Pozzetti et al. 2003]{pozzetti03}
Pozzetti, L., et al. 2003, A\&A, 402, 837
\bibitem[Scannapieco et al. 2005]{scannapieco05}
Scannapieco, E., Silk, J., \& Bouwens, R., 2005, ApJL, 635, 13
\bibitem[Sijacki, springel]{Sijacki05}
Sijacki, D., \& Springel, V., 2006, MNRAS, 366, 397
\bibitem [Silk 2005]{silk05}
Silk, J. 2005, MNRAS, 364, 1337
\bibitem[Somerville  99]{somerville99}
Somerville, R. S., \& Primack, J. R., 1999, MNRAS, 310, 1087
\bibitem[Somerville 2004a]{somerville04a}
Somerville R. S., 2004a, in Bender R., Renzini A., eds, Proc.ESO/USM/MPE Workshop, 
Multiwavelength Mapping of Galaxy Formation and Evolution. Springel-Verlag, Heidelberg (astro-ph/0401570)
\bibitem[Somerville et al. 2004b]{somerville04b}
Somerville, R. S., et al., 2004b, ApJ, 600, L135
\bibitem[Somerville et al. 2004c]{somerville04c}
Somerville, R. S., Lee, K., Ferguson, H. C., Gardner, J. P., 
Moustakas, L. A., \& Giavalisco, M., 2004c, ApJ, 600L, 171  
\bibitem[van Dokkum et al. 2006]{van Dokkum06}
van Dokkum, P. G., et al. 2006, ApJL in press (astroph/0601113)
\bibitem[Yan \& Thompson]{yan03}
Yan, L., \& Thompson, D., 2003, ApJ, 586, 765
\bibitem[Yan et al. 2004]{yan04}
Yan, L., Thompson, D., Soifer, B. T., 2004, AJ, 127, 1274
\end{thebibliography}
\end{document}